\documentclass[12pt,a4paper,final]{iopart}

\usepackage{iopams}  
\expandafter\let\csname equation*\endcsname\relax
\expandafter\let\csname endequation*\endcsname\relax
\usepackage{amsmath}

\usepackage{graphicx}
\usepackage[breaklinks=true,colorlinks=true,linkcolor=blue,urlcolor=blue,citecolor=blue]{hyperref}
\usepackage{braket}
\usepackage{subfig}

\begin{document}

\title[]{Propagating wave correlations in complex systems }

\author{Stephen C Creagh, Gabriele Gradoni, Timo Hartmann and  Gregor Tanner}
\address{School of Mathematical Sciences, University of Nottingham, UK}
\eads{\mailto{gregor.tanner@nottingham.ac.uk}}

\begin{abstract}
We describe a novel approach for computing wave correlation functions
inside finite spatial domains driven by complex and statistical sources.
By exploiting semiclassical approximations, we provide explicit 
algorithms to calculate the local mean of  these correlation functions in 
terms of the underlying classical dynamics. 
By defining appropriate ensemble averages, we show that fluctuations about the mean 
can be characterised in terms of classical correlations.
We give in particular an explicit expression relating fluctuations of diagonal contributions to those of 
the full wave correlation function. 
The methods have a wide range of applications both in quantum mechanics and for 
classical wave problems such as in vibro-acoustics and electromagnetism. 
We apply the methods here to simple quantum systems, so-called quantum
maps, which model the behaviour of generic problems on Poincar\'e sections. 
Although low-dimensional, these models exhibit a chaotic classical limit and share
common characteristics with wave propagation in complex 
structures. 
\end{abstract}
\vspace{2pc}
\submitto{\JPA}

\section{Introduction}

There is a long tradition of describing statistical properties of wave fields and 
spectra in terms of semiclassical or high-frequency techniques and random 
matrix arguments, both for classical waves such as vibroacoustics
\cite{TS07, WW10}, electromagnetics \cite{GG14, Gro14} 
and for quantum mechanics \cite{GMW98, Sto99, Haa01, Meh04}. Spectral and wave function statistics are universal 
under very general conditions and, in particular, if the system under consideration is complex with an associated 
ray dynamics being chaotic \cite{BGS84}. The statistical properties depend then 
only on a few generic parameters such as the 
mean level density, the mean wave amplitude and underlying symmetries such as  time-reversal symmetry. Deviations 
from universality may occur due to non-chaotic dynamical features and 
for many wave systems in practical engineering applications, these deviations are of great practical importance. 
In addition, the spatial dependence of mean field values is itself of interest, 
particularly if absorption or other dissipative effects become relevant. 

Classical wave problems -- such as the vibro-acoustic response of complex 
built-up structures or the electromagnetic wave field inside a complex
cavity -- have in common that they are  
typically driven by complex, that is, spatially extended but finite, stochastic, and broad-band sources.
In addition, they are typically dissipative and are often composed both of regular components 
(rectangular rooms, walls or corridors, for example) and irregular components 
(such as cable bundles,  circuit boards, moulded vehicle parts  or support beams) whose exact  
location, shape, and topology may be uncertain.  Such problems 
pose computational challenges for full numerical
simulation, particularly in the high frequency regime where the 
scale of a wavelength is small compared to the size  of
the structure. The inherent uncertainty that one commonly
encounters in the intrinsic geometry may in any case make detailed 
characterisation of the response irrelevant, even if it was computable.

We therefore take an approach in this paper which uses ray
propagation to predict averaged, coarse-grained features of the response, as well
as providing statistical information. The averages are not sensitive to structural changes, while
providing a platform for the prediction of more detailed 
statistical characteristics in a post-processing step. The connection
between ray propagation and wave-field correlations is based here on 
transfer or boundary operator approaches \cite{CHT13}. Transfer
operators allow us to present the problem of wave propagation via
multiple reflection in a format which mirrors very closely the
phase space mappings used for ray propagation (while in principle
allowing an exact solution). This transfer operator approach means
that we effectively characterise the problem in terms of a
Poincar\'e-section representation, but we emphasise this 
lower-dimensional representation can ultimately be mapped to the full 
problem by using Green function identities (see \cite{GCT15}) for a
detailed discussion of propagation of a correlation function from a 
straight boundary in this context). 

Furthermore, an explicit quantitative 
connection with phase space densities can be made by presenting the 
correlation function as a Wigner function.
The Wigner distribution function (WDF) approach has been developed in the context of quantum mechanics \cite{Hillery1984}, but has found widespread 
applications also for microwaves \cite{Sto99} and in optics \cite{Dragoman_1997,torre2005,Alonso_2011}. 
The method introduced below exploits a connection between the 
field-field \emph{correlation function} (CF) and the WDF \cite{Littlejohn_1993,Winston_2006,Dittrich_2009}.
Both quantities have been studied intensively in the 
physics and optics literature. For wave chaotic systems, Berry's conjecture postulates
a universal CF equivalent to correlations in Gaussian random fields
\cite{Berry01JPA_1977, Hemmady2012,GG14}. 
Non-universal corrections can be
retrieved by making a link between the CF and the Green function of the system after suitable averaging
\cite{Creagh_1997,Hortikar_1998,Weaver_2001,Urbina_2006}. 
The Wigner function formalism can be used to derive the ray-tracing limit for 
propagating wave fields, see for example \cite{GCT15}. Higher-order (Airy-function) 
correction to a ray-tracing approach as well as treating evanescent contributions have been discussed in 
\cite{Marcuvitz_1991,Littlejohn_1993,GCT15}. 

In this paper, we will use the WDF approach inside finite domains which may be regularly or irregularly shaped. 
Extending the treatment in \cite{GCT15}, we will consider here multiple reflections and interference effects explicitly.
We derive the ray-tracing limit in this case and discuss various limits leading to deviations
from a pure ray-tracing approach. The ray-tracing component itself can be evaluated 
efficiently when combined with fast phase-space propagation methods such the 
{\em Dynamical Energy Analysis} (DEA) developed in the context of vibro-acoustics \cite{Tanner2009} and 
mesh-based implementation tools such as {\em Discrete Flow Mapping} (DFM) techniques \cite{Chappell2013}.

We will focus here on linear and stationary (frequency domain) scalar wave 
problems using as reference point the wavenumber $k$ or the frequency $\omega$. 
When considering quantum systems, we may identify the scale 
$1/k$ with $\hbar$; extensions to vector wave equations is straightforward \cite{GG15}. 
In Sec.\ \ref{sec:trans}, we introduce the concept of transfer operators and
write correlation functions of stationary wave fields in terms of these operators. We derive
relations between correlation functions and phase space densities and verify the 
results with the help of a simple quantum map, the kicked Harper map \cite{Art_Scholar}. 
In Sec.\ \ref{sec:var}, we extend the results to derive explicit expressions for the variance around 
the ensemble mean in terms of classical phase space quantities. The results are again validated 
numerically, here for a fully chaotic quantum map, the perturbed cat map \cite{SCC94}.

\section{Transfer operator formulation and correlation functions} \label{sec:trans}
\subsection{The transfer operator $\hat{T}$}
\begin{figure}[htb!]
 \centering
 \includegraphics[scale=0.30]{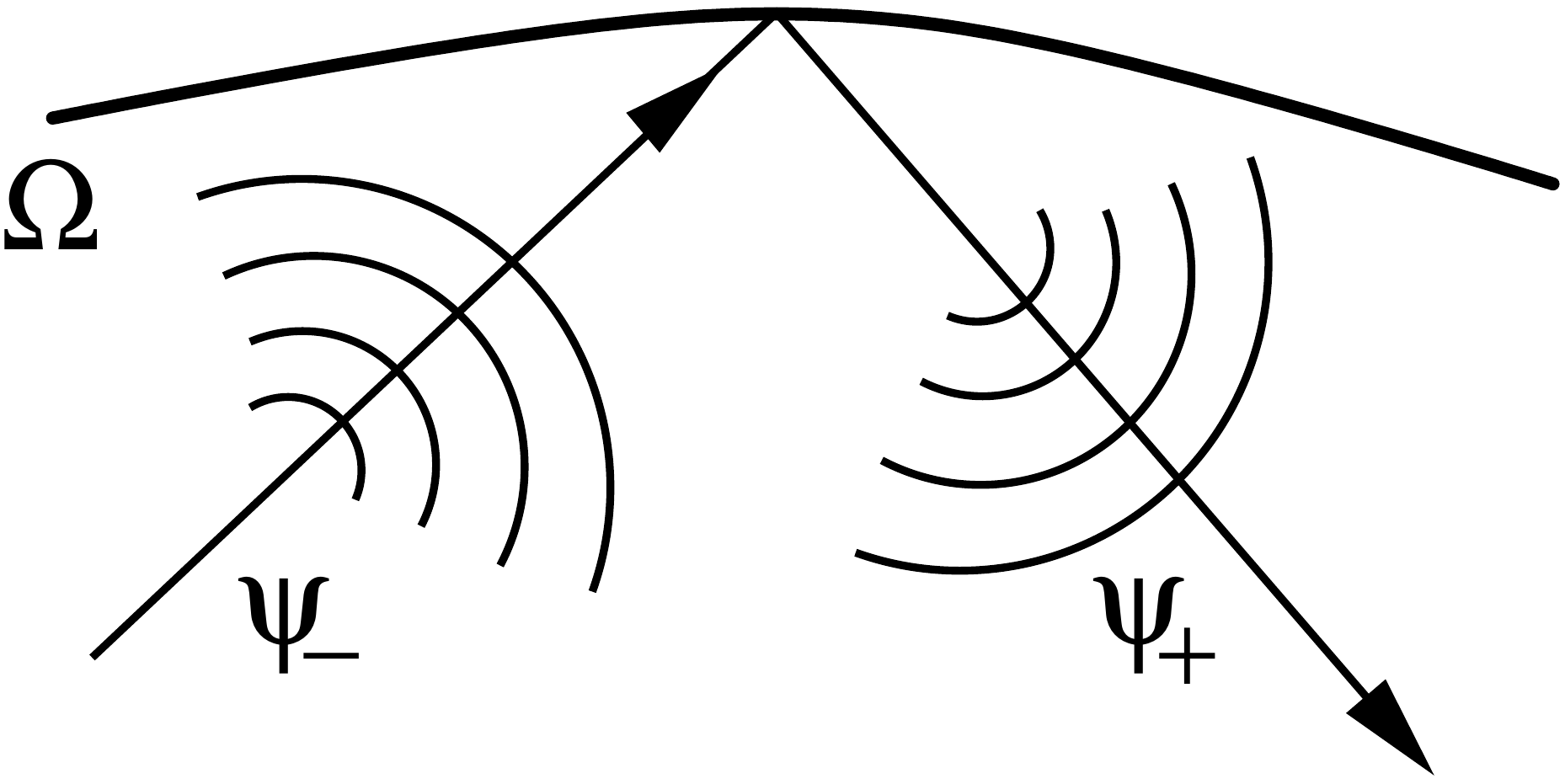}
 \caption{\label{sec:inout} Boundary field decomposed into incoming and outgoing component. }
\end{figure}

The wave problem under consideration is described in an operator
formulation using transfer operators $\hat{T}$ defined on a
$(d-1)$-dimensional manifold, or surface of section (SOS), where 
$d$ is the dimension of the underlying space. As a concrete example,
we consider the Helmholtz equation in a  $d$-dimensional region 
$\Omega \subset \mathbb{R}^d$ 
\begin{equation} \label{helm}
-\nabla^2\psi - k^2\psi = \psi_0.
\end{equation}
Boundary conditions given on $\partial \Omega$ are assumed to take the form of a linear relationship
between the solution $\psi$ itself and its normal derivative 
$\partial \psi/ \partial n$. We will in general use $\partial \Omega$ 
as the SOS although other choices are possible. 
We include an inhomogeneous source term $\psi_0$ here, which drives
the wave dynamics.  An operator formulation 
on the boundary is naturally given in terms of boundary integral equations, which yield
relations between the wave function $\psi$ and its normal derivative
on the boundary. 
For our purposes, it is more convenient to follow
\cite{CHT13} and decompose the  boundary field into incoming and outgoing components
\begin{equation} \label{psim}
\ket{\psi} = \ket{\psi_-} + \ket{\psi_+}
\end{equation}
as depicted in Fig.\ \ref{sec:inout}. Note that we use the ket notation
for the solutions restricted to the SOS given here by $\partial \Omega$.

The solution to the inhomogeneous wave problem can now be cast in
terms of incoming waves at boundaries, 
\begin{equation} \label{prop}
(1 - \hat{T}) \ket{\psi_-} = \ket{\psi_-^0},
\end{equation}
where $\ket{\psi_-^0}$ is the source wave field incoming on the boundary. 
Transfer operators can be defined for generic wave systems and for generalised SOSs other 
than $\partial \Omega$, see \cite{Bog92, TS07a}. 
An explicit construction of the exact $\hat{T}$ is, however, often not straightforward and implementations have 
been presented for model systems only \cite{Pro94, RS95,Pro96,CHT13}. 
General expressions for $\hat{T}$ can be given explicitly in the 
semiclassical limit $k$ (or $\omega$)  $\to \infty$ in terms of the so-called 
Bogomolny transfer operator \cite{Bog92, Dor92},
\begin{equation} \label{BoT}
T({x},{x}';\omega) \approx \left(\frac{k}{2\pi \rmi}\right)^{(d-1)/2}
\sum_{\stackrel{{\rm rays}}{{x}'\to {x}}} 
D(x,x') \exp\left(\rmi k S({x},{x}';\omega) - \rmi \mu
\frac{\pi}{2} \right)
\end{equation}
with 
\begin{equation} \label{BoT1}
D(x,x') = \sqrt{\left|\det \left(\frac{\partial^2 S}{\partial x \partial x'}\right)
\right|} .
\end{equation}
The sum in (\ref{BoT}) is over all rays passing through the 
interior to cross from point $x'$ to point $x$ on the $(d-1)$ dimensional
SOS. The phase $S(x,x')$
is the classical action at a fixed frequency $\omega$ of the ray and 
$\mu$ is a phase index arising due to boundary
conditions or ray caustics; $S$ can be complex
 if there is damping. Note that, when considering the Helmholtz case
 in a domain $\Omega$, there is only one such ray trajectory from $x'
 \to x$. 
The wave number $k$ is introduced here for convenience; in the special 
case of the Helmholtz equation (\ref{helm}), we have $S(x,x') \equiv k
\, L(x,x')$, the optical  length of the chord connecting  $x'$ to  $x$. 

\subsection{The correlation function $\hat{\Gamma}$} 
\label{sec:corr}

We are primarily interested in classical wave problems driven by
noisy, stochastic sources. 
In this context, the natural solution of the
problem is in terms of a two-point correlation function
\[ \Gamma(x_1,x_2) = \bra{x_1}\hat{\Gamma}\ket{x_2}, \]
where $\hat{\Gamma}$ can be interpreted as a density matrix of the
system and $x_1$ and $x_2$ are coordinates on the SOS. 
In the simplest case of coherent driving, we may consider 
$\hat{\Gamma} = \ket{\psi^-}\bra{\psi^-}$, with $\ket{\psi^-}$ 
a solution of the form (\ref{psim}) for a specific source term and 
at a given frequency.  More generally, we consider driving by
stochastic sources and correspondingly
density matrices obtained from, for example,  
frequency, spatial or time averages (when considering time dependent, 
in particular, non-harmonic, stochastic sources) \cite{EMC15}. 
Note that we always assume here that $\hat{T}$ itself is not explicitly time dependent, even if the
driving is. 

Consider a source correlation function $\Gamma_0$ with
\begin{equation} \label{Gx1x2}
\Gamma_0(x_1,x_2) = \braket{\psi_0^-(x_1) \psi^-_0(x_2)^*},
\end{equation}
where $\psi_0^-(x)$ is the source wave field as defined in (\ref{psim}) and $\braket{.}$ denotes an ensemble average over time intervals, 
frequency or local spatial averaging. Note, that while the source distribution $\ket{\psi_0^-}$ show strong spatial 
fluctuations, the averaged quantity ${\Gamma_0}$ can be a fairly smooth function of both $x_1$ and $x_2$.
Starting from (\ref{prop}), we write the system correlation function, including multiple reflections at boundaries, in terms of the source correlation function and the 
transfer operator $\hat{T}$. That is,
\begin{equation} \label{corr-def}
\hat{\Gamma}  = (1 - \hat{T})^{-1} \hat{\Gamma}_0 (1 - \hat{T}^{\dagger})^{-1}.
\end{equation}
Formally, one can write this in terms of the source correlation
function $\Gamma_0$ being propagated along outgoing waves and
undergoing multiple reflections at boundaries. After geometrically expanding the right hand side of Eq.\ (\ref{corr-def}), we write
\begin{equation} \label{G0}
\hat{\Gamma}  = \sum_{n,n' = 0}^{\infty} \hat{T}^{n} \hat{\Gamma}_0 (\hat{T}^{\dagger})^{n'} = \hat{K} + 
\sum_{m=1}^{\infty} \left(\hat{T}^m \hat{K} + \hat{K} (\hat{T}^{\dagger})^m\right), 
\end{equation}
where the $n$th iteration of the operator $\hat{T}^n$ is related to
waves undergoing $n$ reflections at $\partial \Omega$.  
The operator $\hat{K} $ in (\ref{G0}) contains the diagonal contribution of the double sum and is defined as
\begin{eqnarray} \label{K}
\hat{K} = \sum_{n=0}^\infty \hat{K}_n &=& \sum_{n=0}^{\infty}  \hat{T}^{n} \hat{\Gamma}_0 (\hat{T}^{\dagger})^{n} .
\end{eqnarray}
It contains the smooth part of the correlation function as will be shown below. The terms 
$\hat{K}_n$ can be interpreted as the $n$th iteration or reflection
contribution to the smooth part of the correlation function. Note that some damping is
implicitly assumed here in order for these sums to converge.

We finally remark that, once the correlation function has been
characterised in a boundary representation, as set out in this section,
one can use Green function identities to propagate this information to
the interior. See \cite{GCT15} for a more detailed discussion of such 
propagation of correlation functions from a straight boundary,
including the treatment of evanescent components, for example.

\subsection{Relation to classical phase space densities} 
The quantities $\hat\Gamma_0$, $\hat\Gamma$ and $\hat{K}$ can be 
related to phase space densities using Wigner transformation 
after additional averaging as demonstrated below and in  \ref{app:A}. 
The WDF of an operator $\hat\Gamma$ is defined as 
\begin{equation} \label{WDF}
W_{\Gamma}(x,p) 
= \int {\rm d}s \, \e^{-\rmi k p s}\, \Gamma\left(x+\frac{s}{2}, x-\frac{s}{2}\right)\, ,
\end{equation} 
with back transformation given by 
\begin{equation}\label{WDFback}
  \Gamma(x_1,x_2) =  \left ( \frac{k}{2 \pi} \right )^{d-1} \int {\rm
    d}p \, 
\e^{\rmi k (x_1-x_2)p} \; W_\Gamma \left(\frac{x_1+x_2}{2}, p \right).
\end{equation}
Here, $k$ represents a wave number such as introduced in (\ref{helm}),
(\ref{BoT}) and $d$ is the dimension of the full, interior problem. It is shown in 
\ref{app:A}, that for sufficiently smooth initial phase space
distributions $W_{\Gamma_0} = \rho_0$, the averaged Wigner transform of 
$\hat{K}_n$ in  (\ref{K}) is given in terms of the classical flow equations,
see also \cite{GCT15,MWH01}. 

Classical phase space densities are driven by the phase space dynamics, in our case a boundary map or 
more generally the Poincar\'e map of an SOS. The initial density $\rho_0 = W_{\Gamma_0}$ can then be associated 
with a boundary density of rays arriving directly from a source distribution in the interior. Mapping the 
source ray density through subsequent reflections leads to the iterated densities
$\rho_0\to\rho_1\to\cdots\rho_n\to\cdots$
which can be described in terms of the (linear) integral operator
 $\cal L$ defined in the lossless limit as \cite{book}
\begin{equation} \label{FP}
{\cal L} [\rho_n](X) = \rho_{n-1}(\varphi^{-1}(X)) =  
\int {\rm d} X' \delta(X-\varphi(X'))\rho_{n-1}(X')
\end{equation}
(see below for the treatment of losses).
Here, $X=(x,p)$ denotes the collective phase space coordinates on the
SOS (with $p$ the momentum conjugate to $x$) and $\varphi:X'\to X$ is the classical map describing the flow of 
trajectories from the SOS back to itself after a single reflection.
We use that $\varphi$ is Hamiltonian, and, in particular, 
phase space volume preserving. The operator
$\cal L$ is also referred to as the Frobenius-Perron (FP) operator
\cite{book}. The integral representation in (\ref{FP}) is useful for considering effects 
like absorption and mode conversion as well as uncertainty, see \cite{CT14}. 
The ray-dynamical, or classical, analogues of equations (\ref{K}) are now provided by
\begin{equation}\label{classprop}
\rho = {\cal L}\rho + \rho_0
\quad\Rightarrow\quad \rho=\frac{1}{1-{\cal L}}\, \rho_0=
\sum_{n=0}^\infty {\cal L}^n\, \rho_0
= \sum_{n=0}^\infty \rho_n.
\end{equation}
From the relation derived in \ref{app:A}, we can now write
\begin{equation} \label{Wrho}
\langle W_{K_n}(X)\rangle \approx 
W_{\Gamma_0}\left(\varphi^{-n}(X)\right)
= {\cal L}^n \left[ \rho_0 \right](X) = \rho_n(X), 
\end{equation}
and we obtain for the Wigner function of the full $\hat{K}$ operator to leading order 
\begin{equation} \label{FPK}
\langle W_K \rangle = \sum_{n=0}^\infty \langle W_{K_n} \rangle\approx \sum^{\infty}_{n=0} {\cal L}^n[\rho_0] = \frac{1}{1-{\cal L}}\,  \rho_0 = \rho\, .
\end{equation}
The averaging $\langle \cdot \rangle$ is understood as defined in (\ref{Gx1x2}) and can be performed in terms of an 
average over an ensemble of similar systems, appropriately chosen frequency averaging or local  (spatial) averaging, for example.
In addition, it is assumed that the initial density $\rho_0 = W_{\Gamma_0}$ is a smooth function on the scale of $\Delta x \Delta k = 1$.

For the full correlation function, we find that contributions to (\ref{G0}) with $n\ne n'$ are removed by 
averaging so that we may assert in addition that
\begin{equation} \label{mean}
\langle W_\Gamma\rangle=\langle W_K\rangle \approx \rho.
\end{equation}
The quantities $\hat{\Gamma}$ and $\hat{K}$ thus have the same mean when taking a suitable average and this mean value is given 
by the classical equilibrium (phase space) density $\rho$ obtained from (\ref{FPK}), that is, 
\begin{equation} \label{avGK}
\langle \hat{\Gamma}\rangle = \langle \hat{K} \rangle \approx W^{-1}[\rho].
\end{equation}
We will find in the next section, however, that fluctuations of $\hat\Gamma$ about this mean are much stronger than those of
$\hat{K}$, particularly in the limit of weak damping. 

The averaged quantities in (\ref{avGK}) contain detailed information 
about the system. Powerful numerical tools have been developed for  
computing $\rho$, and thus indirectly for the mean of $\hat{\Gamma}$. 
Among those, the DEA technique together with the DFM implementation 
on meshes is particularly suitable for complex structures, which has been used in 
a range of engineering applications
\cite{Tanner2009, Chappell2013, CLST14}. 

Note that, without losses, the FP operator $\cal L$ has a leading eigenvalue 1 and the solution to 
Eq.\ (\ref{FPK}) diverges. It is thus necessary - in order to arrive at
mathematically and physically meaningful solutions - to account for losses
such as naturally
occur in wave systems due to 
wall absorption or absorption in the interior of $\Omega$ or due to
radiation through apertures. 
Absorption can be included in a FP operator formalism by introducing, for example,      
\begin{eqnarray} \label{FP_damp}
{\cal L}_\mu[\rho_0](X) &=& \int e^{-\mu(X')} \delta\left(X - \varphi(X')\right) \rho_0(X')\\ 
&=& e^{-\mu(\varphi^{-1}(X))} \rho_0[\varphi^{-1}(X)],
\end{eqnarray}
where we require that the damping coefficient $\mu \ge 0$ is additive along the
flow, mimicking distributed losses in many real-life engineering systems. 
For interior damping, such as for acoustic waves in air, we typically have  $\mu(X) = \mu_0\,  L(X)$, 
where $L$ is the length of a trajectory segment between two reflections on the boundary starting at $X$ 
and $\mu_0\ge 0 $ is a real constant.

\subsection{Numerical illustration of averaged response using quantum map models}\label{sec:QKH}
We will demonstrate the validity of the relations described in the
last section  with the help of a simple model in which the transfer
operator is simulated by
\[
\hat{T} = v \hat{U},
\]
where $\hat{U}$ is a unitary operator acting on a space of
dimension $N$ and the prefactor $v$, satisfying $0<v<1$, accounts for dissipation. In
numerical investigations here and in following sections, we choose $\hat{U}$ to be the 
quantum analogue of a classical map on the unit torus (see \ref{app:B} for
further details). By choosing maps with different levels of chaotic behaviour, we aim to simulate the response of
larger, complex systems governed by classical wave theories. In these models, 
the dimension $N$ takes on the
role of the wave number $k$ as the large parameter. 

\begin{figure}[tb]
\centering
  \vspace{-0.9cm}
  \captionsetup[subfloat]{captionskip=-0.0em,singlelinecheck=off,justification=raggedright}
  \subfloat[]{\label{fig:kr:wig:a0}\includegraphics[width=7.2cm]{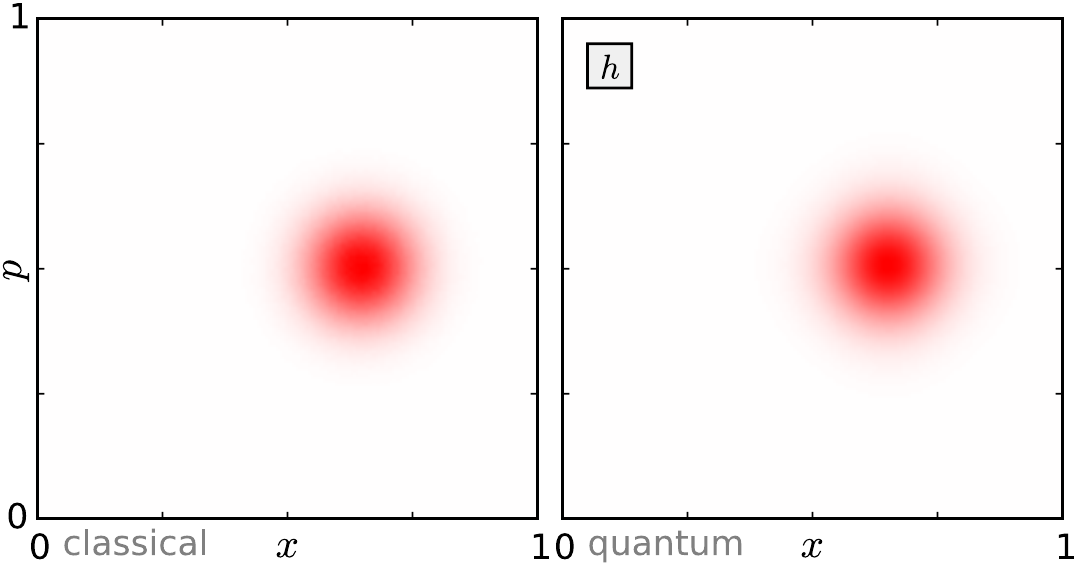}}\hspace{.1cm}
  \subfloat[]{\label{fig:kr:wig:a1}\includegraphics[width=7.2cm]{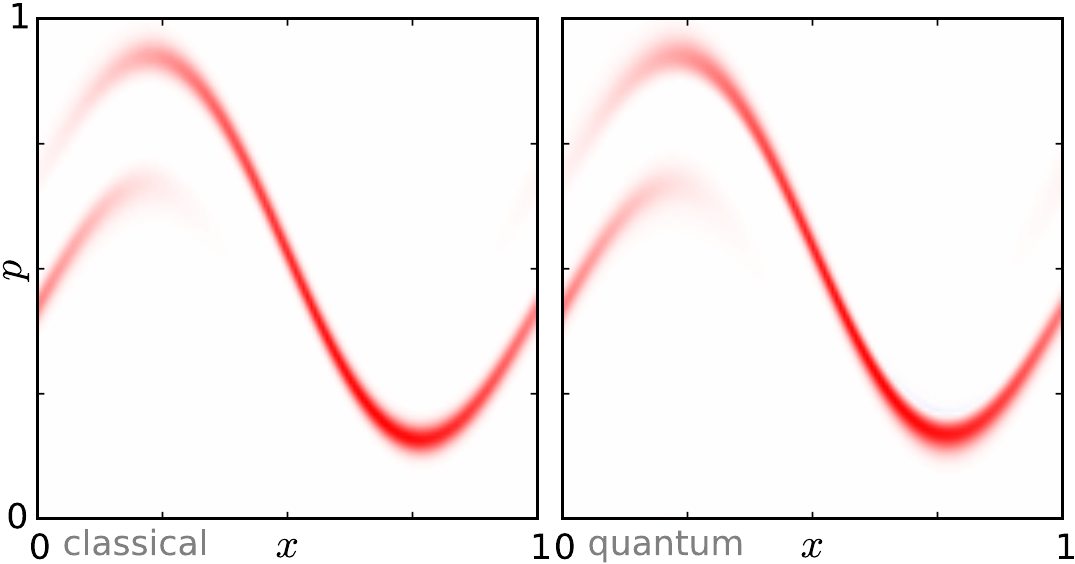}}\\[0.0cm]
  \subfloat[]{\label{fig:kr:wig:a3}\includegraphics[width=7.2cm]{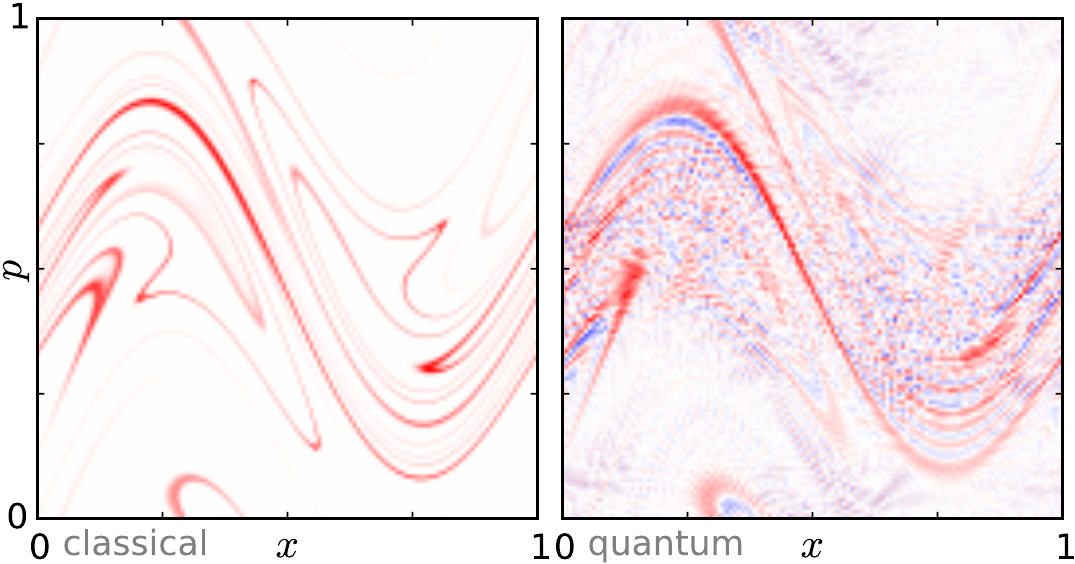}}\hspace{.1cm}
  \subfloat[]{\label{fig:kr:wig:a5}\includegraphics[width=7.2cm]{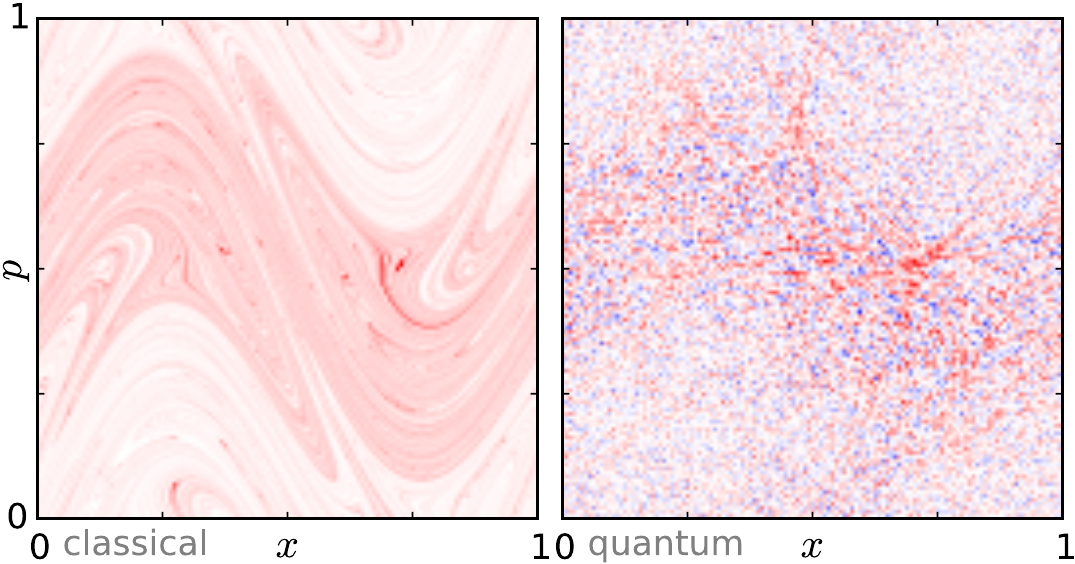}}
  \caption{ \label{fig:kr:wig:a}
     The phase space dynamics of the classical and quantum kicked Harper map for $N = 128$ and without damping ($v=1$); 
     the initial distribution is chosen according to Eq.~(\ref{WG0}) with 
     $x_0\mathord{=}0.65$, $p_0\mathord{=}0.5$ and $\sigma_x\mathord{=}\sigma_p \mathord{=} 0.075$.
     The left panel shows the classical density $\rho_n$, where $n$ is the number of iterations 
     (using $10^7$ particles). The
     right panel shows the quantum Wigner function $W_{K_n}$ on a $128\times 128$ grid with
     \protect\subref{fig:kr:wig:a0} $n=0$,
     \protect\subref{fig:kr:wig:a1} $n=1$,
      \protect\subref{fig:kr:wig:a3} $n=3$,
      \protect\subref{fig:kr:wig:a5} $n=10$ kicks. The blue regions
      denote negative values of the Wigner function. 
A box of area $h=1/N$ has been included in part (a) in order to give a
visual sense of the quantum scale.
     }
\end{figure}

In order to illustrate the average response set out in the preceding
discussion, we let $\hat{U}$ take the form of a 
 quantum kicked Harper (QKH) map in this
 section,  see \cite{LKFA90, GKP91, LS91} or the review  \cite{Art_Scholar} (other
 maps are used when discussing fluctuations about the mean later). The
corresponding classical kicked Harper map  exhibits a 
predominantly chaotic dynamics but with small regular
islands. Although these islands are small, their presence leads to a
significant degree of tanglement of the stable and unstable manifolds as demonstrated in Figs.\ 
\ref{fig:kr:wig:a3} and \ref{fig:kr:wig:a5}.
This gives rise to a strong dependence on initial source distributions when considering correlation functions, which makes
 it a good test model for studying relations between classical and wave operators. 

The source $\hat{\Gamma}_0$ is chosen  
so that the associated Wigner distribution function takes the form
\begin{equation} \label{WG0} 
W_{\Gamma_0}(x,p)= C \e^{-(1-\cos 2\pi(x-x_0))/(2\pi\sigma_x)^2-(1-\cos 2\pi(p-p_0))/(2\pi \sigma_p)^2}, 
\end{equation} 
where $C$ is a normalisation constant chosen so that $\Tr\Gamma_0=1$
and $\sigma_x$ and $\sigma_p$ respectively determine the variances in
the $x$ and $p$ coordinates. For small enough $\sigma_x$ and
$\sigma_p$ this is a nearly Gaussian source distribution in phase
space centred on $X_0 = (x_0,p_0)$, and is associated with a
corresponding nearly Gaussian source correlation function 
$\Gamma_0(x_1,x_2)$.

The time evolution of the Wigner transform $W_{K_n}$  of
$\hat{K}_n$ is displayed in Fig.\ \ref{fig:kr:wig:a}  and is compared with the evolution of a classical phase space 
density with initial distribution according to (\ref{WG0}). The quantum evolution follows the classical 
dynamics closely for small $n$ as expected from Eq.\ (\ref{FP}) and detailed
in \ref{app:A}. Once the classical dynamics foliates phase space down
to phase space cells of area $1/N$, wave fluctuations take over and wash out the classical phase space
structures, see Fig.\ \ref{fig:kr:wig:a5} at $n=10$. Note that, after
several interations of the map, the classical density is stretched
along the unstable manifolds of the system and tracks, for example, the
folds and switchbacks in them that arise from the small islands present. Some
of these features are also seen in the propagated Wigner function but
the details are lost. 

\begin{figure}[tb]
\centering
  \vspace{-0.7cm}
  \captionsetup[subfloat]{captionskip=0em,singlelinecheck=off,justification=raggedright}
  \subfloat[]{\label{fig:kr:k1:0}\includegraphics[width=0.8999\linewidth]{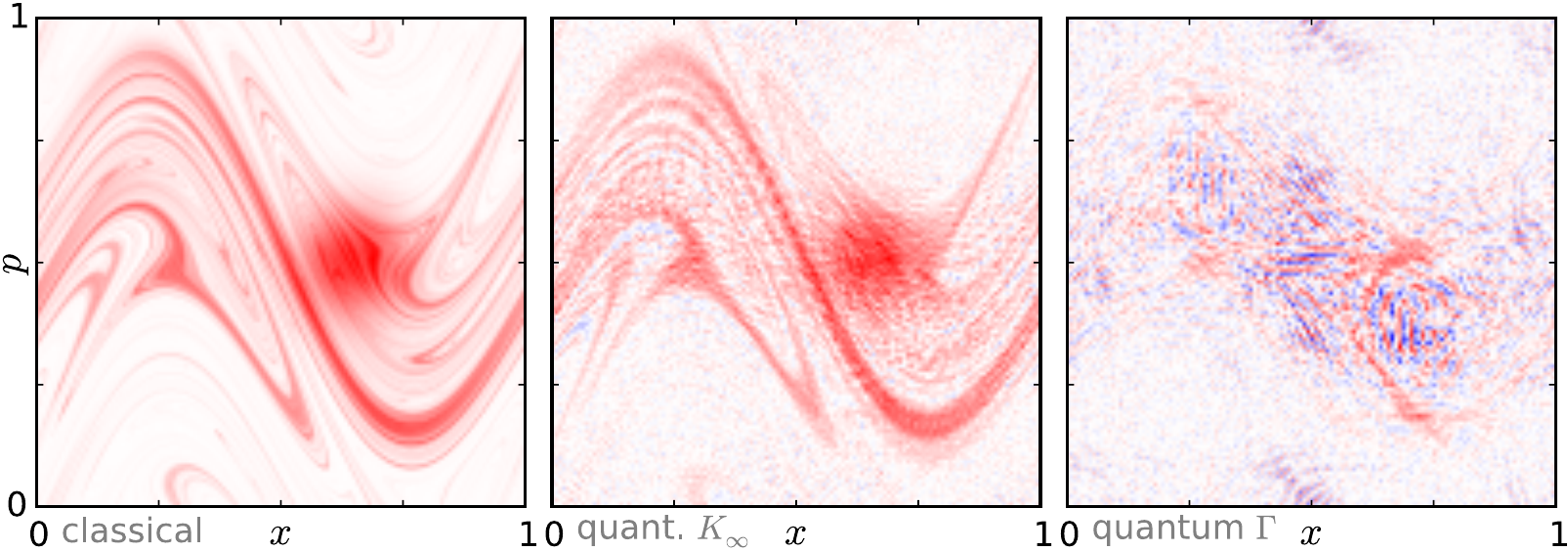}}\\[-.0cm]

  \subfloat[]{\label{fig:kr:k1:2}\includegraphics[width=0.8999\linewidth]{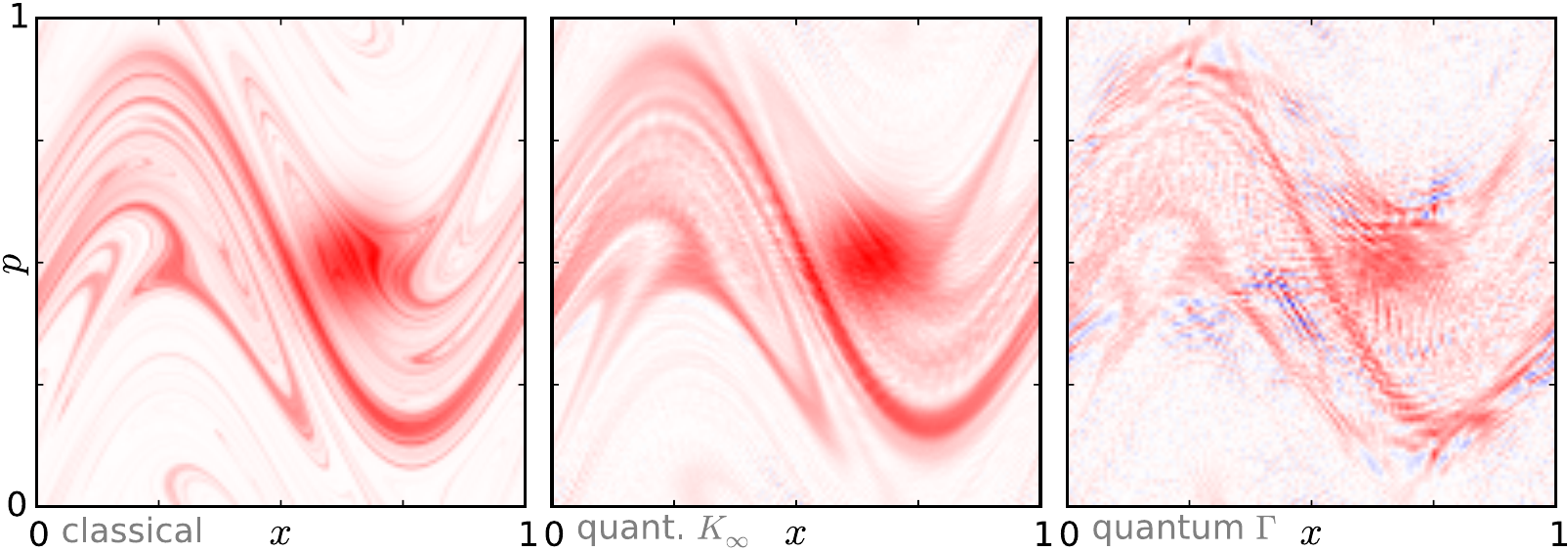}}%
   \caption{ \label{fig:kr:k1:a} The stationary phase space density $\rho$ (left) is compared with the quantum distributions $W_{K}$ (middle panel) and with the 
  Wigner function of the full stationary correlation function $W_{\Gamma}$ (right) for $N=128 $ and $v=0.9$. The parameter for the initial distribution are 
       $x_0\mathord{=}0.65$, $p_0\mathord{=}0.5$ and $\sigma_x\mathord{=}\sigma_p \mathord{=} 0.075$.
     In \protect\subref{fig:kr:k1:2},
     an average over $15^2$ different values of the Harper map parameters (in a range of $\pm 5\%$) is performed in the quantum version but not in the classical version; no averaging has been done in  \protect\subref{fig:kr:k1:0} .
     }
\end{figure}

In Fig.\ \ref{fig:kr:k1:a}, we compare the classical (stationary) phase space distribution $\rho$
defined in (\ref{classprop}) with the Wigner transforms of $\hat{K}$ and $\hat{\Gamma}$ for a 
damping value $v = 0.9$.  The classical and wave results coincide remarkably well for $\rho$ and $W_{K}$
even without further averaging, see Fig.\ \ref{fig:kr:k1:0}, while only traces of the classical phase space structure 
are left for $W_{\Gamma}$ in this case. When performing a further parameter average in the Harper map, the quantum
fluctuations are suppressed and the classical phase space structure comes out more clearly, both for $W_{K}$ and $W_{\Gamma}$, see
Fig.\ \ref{fig:kr:k1:2}. The fluctuations in $W_{\Gamma}$ are in both cases much larger than in $W_{K}$, with similar differences to be 
expected for the original objects $\hat{\Gamma}$ and $\hat{K}$. We can in fact quantify the increased fluctuations in 
$\hat{\Gamma}$ compared to $\hat{K}$ as a function of $v$, as will be described in the next section.

The results show that the classical phase space density gives the mean value for both quantities $W_{K}$ and $W_{\Gamma}$ as expressed in 
(\ref{mean}). It also shows how well the classical dynamics describes a wave operator such as $W_{K}$ down to a detailed description of the 
complex phase space structure present in a mixed system like the
kicked Harper map. Note, that the wave fluctuations due to higher 
iterates of the map are suppressed here due to the 10\% damping ($v=0.9$) introduced.

\section{Fluctuations of $\hat{K}$ and $\hat{\Gamma}$ about the mean} \label{sec:var}
Although $\hat{K}$ and $\hat\Gamma$ have the same average, 
the fluctuations about the mean are much greater for $\hat{\Gamma}$ than 
for $\hat{K}$ as illustrated in Fig.~\ref{fig:kr:k1:a}, an effect which is greatly 
amplified in the weak damping limit.
In this section we quantify this effect for the simple model
$\hat{T}=v\hat{U}$ introduced in
Sec.~\ref{sec:QKH} by calculating mean values for 
$\Tr \hat{\Gamma}^2$ and $ \Tr \hat{K}^2$.
Formally, these also provide us with variances of individual matrix
elements $K_{ij}$ and $\Gamma_{ij}$ in any given basis, using, for example,
\begin{equation}\label{matrixelement}
\langle|\Gamma_{ij}|^2\rangle_N = 
\frac{1}{N^2}
\Tr \hat{\Gamma}^2.
\end{equation}
Here $\langle \cdot \rangle_N$ denotes an average 
over the $N$-dimensional basis on the space on which 
$\hat{T}$ is defined.

Explicit expressions can be given for these quantities, as 
outlined in the remainder of this section. In particular, 
if we average over a parameter such as frequency (or system 
dimension in the quantum map model), we can demonstrate that
fluctuations in $\hat{\Gamma}$ scale simply with fluctuations 
in $\hat{K}$, according to
\begin{equation}\label{GvsK}
\langle \Tr \hat{\Gamma}^2 \rangle = \frac{1+v^2}{1-v^2}
\langle \Tr\hat{K}^2\rangle.
\end{equation} 
This expression is quite general and is equally valid in 
chaotic and integrable limits, for example. It is also true
irrespective of 
whether the system has time reversal symmetry or not. It provides a direct 
quantitative measure of the greater fluctuations seen in 
$\hat{\Gamma}$, relative to those of $\hat{K}$, as $v\to 1$. 
Furthermore, we will see that the fluctuations of $\hat{K}$ itself can be
obtained using information that is readily available from propagation
of a corresponding classical density, as performed in complex
structures using the DEA method, for example.

Using (\ref{matrixelement}), the scaling relationship (\ref{GvsK})
provides an equivalent scaling 
\begin{equation}\label{varGKmatrix}
\langle |\Gamma_{ij}|^2 \rangle_N = \frac{1+v^2}{1-v^2}
\langle |K_{ij}|^2\rangle_N
\end{equation}
of the relative sizes of corresponding matrix elements. It should be
noted that in practice such averages over all matrix elements are
dominated by diagonal elements in the weak damping limit or when
the source intensity is extended over
length scales much larger than a wavelength: however, it can be shown using alternative but
more involved calculations, not reported here, that these results
can also be applied element by element.
Finally, any statement involving such traces can be expressed also as a
relation between Wigner functions. For example, (\ref{GvsK}) can also
be expressed in the form,
\begin{equation}\label{varGKWig}
\langle|W_\Gamma(x,p)|^2\rangle 
=\frac{1+v^2}{1-v^2}
\langle|W_K(x,p)|^2\rangle 
\end{equation}
using standard properties of Wigner functions. This relation
provides an explicit quantitative statement of the qualitative 
observations made in Fig.~\ref{fig:kr:k1:a}.

\subsection{The variance of the diagonal part $\hat{K}$}
We consider $\hat{K}$ first and write Eq.\ (\ref{K}) as 
\[   \hat{K} = \sum_{n=0}^\infty v ^{2n}\, \hat{\Gamma}_n = \sum_{n=0}^\infty v ^{2n}\,  \hat{U}^n \hat{\Gamma}_0 \hat{U}^{-n} \]
with $\hat{\Gamma}_n = v^{-2n} \hat{K}_n$ representing the smooth $n$-step correlation function as in Eq.\ (\ref{K}). 
Using the cyclic permutability of the trace we may write,
\begin{eqnarray}
{\rm Tr} (\hat{K}^2) 
&=& \sum_{n,n'=0}^\infty v^{2(n+n')} {\rm Tr} \left[\hat{U}^{n-n'} \hat{\Gamma}_0 \hat{U}^{n'-n} \hat{\Gamma}_0 \right]\\
&=& \sum_{n,n'=0}^\infty v^{2(n+n')} {A}(n-n') ,
\end{eqnarray}
where
\[ 
{A}(n) = {\rm Tr}\left( \hat{\Gamma}_0 \hat{\Gamma}_n\right)
\]
is the $n$-step return probability.
Note that ${A}(n) = {\rm Tr} \hat{\Gamma}_0 \hat{\Gamma}_n = {\rm Tr} \hat{\Gamma}_n \hat{\Gamma}_0 = {\rm Tr} \hat{\Gamma}_0 \hat{\Gamma}_{-n} = {A}(-n)$.   
After reordering the double sum, this may be written in the form 
\begin{equation} \label{TrK}
{\rm Tr}( \hat{K}^2) = \frac{1}{1- v^4} \sum_{n=-\infty}^\infty v^{2|n|} {A}(n)\, .
\end{equation}
So far the calculation is exact and completely independent of the
classical limit.

The quantity ${A}(n)$ can again be related to the classical phase space dynamics using 
\begin{equation} 
A(n) = {\rm Tr}\left( \hat{\Gamma}_0 \hat{\Gamma}_n\right) = \left(\frac{k}{2 \pi}\right)^{d-1} \int dX \, W_{\Gamma_0} (X) W_{\Gamma_n}(X)\, .
\end{equation} 
We now make the approximation used in (\ref{FP}) that, on averaging,
the fluctuations in $A(n)$  are removed and that it may be approximated by its classical
or ray-dynamical analogue, the classical phase space autocorrelation function
\begin{equation}\label{tcorr}
\langle A(n)\rangle \approx A_{\rm cl}(n) = \int {\rm d}X\;  \rho_n(X)\rho_0(X)\, .
\end{equation}
Here, $\rho_0$ is the correspondingly averaged classical phase space
density of the source and $\rho_n = {\cal L}^n\rho_0$ its
$n$th-reflection iterate. Then
\[
\langle {\rm Tr}( \hat{K}^2)\rangle \approx \frac{1}{1- v^4}
\sum_{n=-\infty}^\infty v^{2|n|} {A}_{\rm cl} (n).
\]
For strongly chaotic systems, the autocorrelation function decays exponentially as 
\begin{equation} \label{corr}
A_{\rm cl}(n) \sim a_0 + a_1\exp(-\gamma_{\rm cor} n),
\end{equation}
where $\gamma_{\rm cor}$ denotes the correlation exponent and $a_0, a_1$ are constants depending on the 
initial density $\rho_0$ ($\e^{-\gamma_{\rm cor}}$ typically being the 2nd
largest eigenvalue of the FP operator.) 
Note that, for strong damping, it suffices to know the
autocorrelation function $A(n)$ for relatively small $n$ in order to
compute $\Tr (\hat{K}^2)$. In this case
one finds that the exact $A(n)$ may be well approximated by its 
ray-dynamical analogue $A_{\rm  cl}(n)$ even without averaging and
that the result above then also holds for individual maps. (Of course, the trace operation itself performs an
average over basis states even if no further averaging is performed.)

\subsection{The variance of the correlation function $\hat{\Gamma}$}
To get an equivalent expression for the variance of the full correlation function of the stationary wave field,  $\hat{\Gamma}$, we start from  Eq.\ (\ref{G0}), giving 
\[   \hat{\Gamma} = \sum_{n_1, n_2=0}^\infty v ^{n_1+n_2}\,  \hat{U}^{n_1} \hat{\Gamma}_0 \hat{U}^{-n_2} \]
from which we obtain (using cyclic permutability of the trace again)
\begin{eqnarray}
{\rm Tr} \,\hat{\Gamma}^2 &=& \sum_{n_1,\ldots,n_4 =0}^\infty v^{n_1+n_2+n_3+n_4} {\rm Tr} \left[\hat{U}^{n_1} \hat{\Gamma}_0 \hat{U}^{-n_2} \hat{U}^{n_3} \hat{\Gamma}_0 \hat{U}^{-n_4}\right]\nonumber \\ \label{qsum}
&=& \sum_{n_1,\ldots,n_4 =0}^\infty v^{n_1+n_2+n_3+n_4} {\rm Tr} \left[\hat{U}^{n_1-n_4} \hat{\Gamma}_0 \hat{U}^{-n_2+n_3} \hat{\Gamma}_0 \right].
\end{eqnarray}
We argue next that, after averaging,
\[
\langle {\rm Tr}\, [\hat{U}^n \hat{\Gamma}_0 \hat{U}^{-n'} \hat{\Gamma}_0 ] \rangle = {\delta}_{nn'} \langle A(n)\rangle
\] 
due to the phase difference that one necessarily finds  in
contributions to $\hat{U}^n$ and $\hat{U}^{n'}$ when $n\ne n'$. 
On average, the sum in (\ref{qsum}) is then dominated by terms with $n_1 -n_4 = n_2-n_3 = n$ and,
after reordering, we may write
\begin{eqnarray}
\langle{\rm Tr} (\hat{\Gamma}^2) \rangle &=& \sum_{n=-\infty}^\infty \sum_{n_2,n_4 =0}^\infty v^{2(n_2+n_4)} v^{2|n|} \langle{A}(n)\rangle\nonumber \\
&=& \frac{1}{(1-v^2)^2} \sum_{n=-\infty}^\infty  v^{2|n|} \langle{A}(n)\rangle \label{TrG}\\ 
& = & \frac{1+v^2}{1-v^2} \langle{\rm Tr} (\hat{K}^2)\rangle,\label{TrG2}
\end{eqnarray}
as previously reported in (\ref{GvsK}). 
Note that this condition
has been derived based on neglecting 
interfering contributions which are wiped out by appropriate 
averaging. We have not made any further assumptions about the
underlying classical dynamics and the relations (\ref{GvsK}) -- (\ref{varGKWig}) are universal, 
independent of whether the system dynamics 
is regular, chaotic or mixed. In particular, the variance of
$\hat{\Gamma}$ always exceeds that of $\hat{K}$, only approaching 
the same value in the limit of strong damping, $v \to 0$.

We note that relation between wave fluctuations (as a function of energy or frequency)
 and classical decay rates is a well established concept and has been discussed 
in the context of fluctuations in scattering cross sections  
\cite{BS88, DSF90} or conductance fluctuations for transport through open 
chaotic cavities \cite{JBS90}, see also \cite{Bar95}. The relevant classical quantity 
is here the classical escape rate $\gamma_{\rm esc}$, which in our setting corresponds to
the rate of dissipation $\gamma_{esc} =  -\log v$. A generalised treatment involving the 
full spectrum of the FP operator is given in \cite{Agam00}.
Our result highlights the influence of higher order eigenvalues of the FP 
operator on the variances in Eqs.\ (\ref{TrK}) and (\ref{TrG})
such as through the classical decay of correlation contributions, Eq.\ (\ref{corr}).   
The main result of this section is, however, the relation (\ref{TrG2})
relating the fluctuations in the 'smooth' part, $\hat{K}$, to the fluctuations in the 
total correlation function. Together with (\ref{avGK}) and
(\ref{tcorr}), we can now relate {\em the first and second moments of the distributions of both 
$\hat{\Gamma}$ and $\hat{K}$ to classical phase space observables} such as the stationary phase 
space density $\rho$ and the classical phase space autocorrelation function
$A_{\rm cl}(n)$. These quantities depend of course on the underlying classical dynamics.  

\subsection{Numerical illustration of fluctuations using quantum map models}\label{numsec}
The fluctuations of $\hat{K}$ and $\hat{\Gamma}$ are now illustrated
using a quantum map model. As asserted in the preceding discussion, 
averages (\ref{TrK}) and (\ref{TrG}) and the corresponding variances
are insensitive to the underlying symmetries and to how chaotic or
integrable the system is, although the detailed distributions are not.

\begin{figure}[tb]
\centering
  \includegraphics[width=0.9\linewidth]{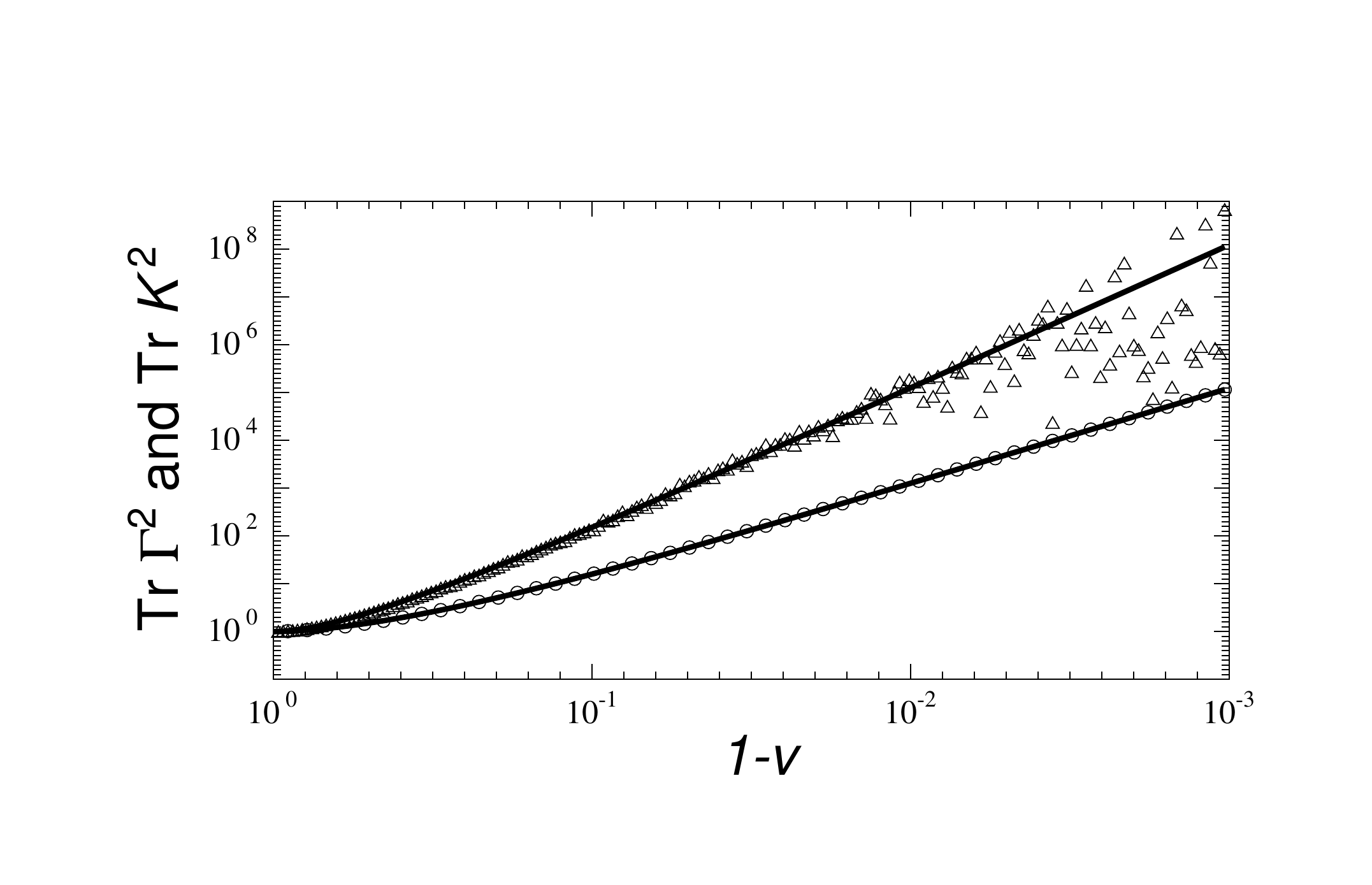}%
   \caption{ \label{fig:fluc} Growth in the fluctuations of $\hat{K}$ and
     $\hat{\Gamma}$ are illustrated as $v\to 1$,  measured respectively by $\Tr
     K^2$ and $\Tr \Gamma ^2$. Calculations are for a nearly Gaussian
     source density $\Gamma_0$ driving a perturbed cat map and centred
     on a fixed point (of period one). Solid
     curves are the averaged predictions (\ref{TrK}) and (\ref{TrG}) based on the classical 
   autocorrelation function $A(n)$. Circles and triangles are
   respectively evaluations of $\Tr K^2$ and $\Tr \Gamma ^2$ for
   individual values of $N$:
   as $v$ is varied, $N$ is stepped from $N=200$ to $N=400$. Note that
   the averages (\ref{TrK}) and (\ref{TrG}) are independent of $N$:
   we let $N$ change in this figure simply to give a sense of the 
   typical variation about the predicted mean for individual maps. This aspect is
   illustrated more explicitly in Fig.~\ref{fig:Gfluc}
   }
\end{figure}

In this section, we use a perturbed cat map \cite{SCC94}, in which a 
quantum cat map \cite{HB80, Kea91, AB95} is perturbed with a QKH map 
of the form used in
Sec.\ \ref{sec:QKH} (see \ref{app:B} for more details). The perturbation is large enough 
to break any underlying time-reversal or spatial symmetries of the
quantum version of the cat map, but small
enough that the overall classical dynamics is still completely chaotic. 
A source term $\Gamma_0$ is used which corresponds to
the nearly Gaussian density given in (\ref{WG0}), with $\sigma_x
= \sigma_p = 1/2$.
It is centred on a period-one fixed 
point $X_0=(x_0,p_0)$ near the origin of phase space, which slows the
short-time decay of the autocorrelation function $A(n)$. 

Corresponding numerically computed 
values of $\Tr \hat{K}^2$ and $\Tr \hat{\Gamma}^2$ are shown as circles and
triangles respectively in Fig.~\ref{fig:fluc} as $v\to 1$. As $v$ is
varied, the dimension $N$ of the quantum map is also changed, 
so that typical variations can be seen in individual values of 
$\Tr \hat{K}^2$ and $\Tr \hat{\Gamma}^2$ around the averages
(\ref{TrK}) and (\ref{TrG}), presented in the figure
as solid curves. Over the range shown, individual values of $\Tr
\hat{K}^2$ follow the average prediction (\ref{TrK}) quite closely.
The trace operation is effectively self-averaging in this case.
In contrast, although (\ref{TrG}) is found to be a good predictor 
of average behaviour, individual values of $\Tr \hat{\Gamma}^2$
are seen to fluctuate significantly about the mean, and to an
increasing extent as $v$ approaches one.

\begin{figure}[tb]
\centering
  \includegraphics[width=0.9\linewidth]{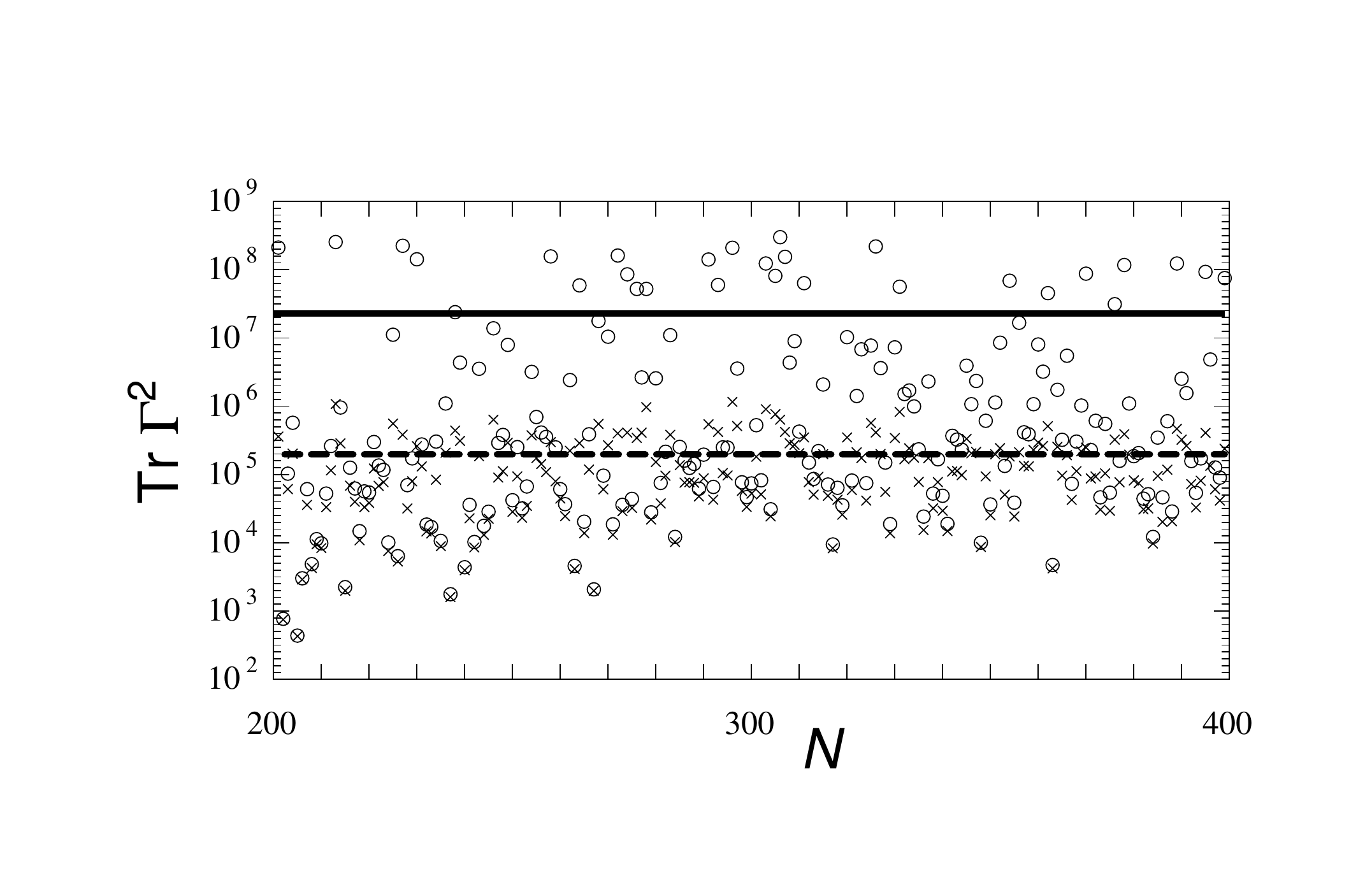}%
   \caption{ \label{fig:Gfluc} Fluctuations in $\Tr \hat{\Gamma}^2$
     are shown as $N$ varies between $200$ and $400$ for  $v=0.995$ and 
     $v=0.999$, represented respectively by
     crosses and circles. The corresponding
     average values are shown as horizontal dashed and solid
     lines, respectively. 
     The smallest values of $\Tr \hat{\Gamma}^2$ are insensitive 
     to the value of $v$ (the crosses are near the circles at the
     bottom of the graph), but the largest values change significantly
     as $v$ approaches unity. 
}
\end{figure}

In fact, it is found in Fig.~\ref{fig:fluc} that most individual 
values of $\Tr \hat{\Gamma}^2$ fall significantly below the mean (\ref{TrG}) 
when $v$ is very close to one, with the average being achieved because
increasingly rare individual cases arise with exceptionally large 
values. This can be understood simply in terms of the physics of
the weakly damped resonant response of the system and is illustrated
in more detail in  Fig.~\ref{fig:Gfluc}. Here we present fluctuations
with changing $N$ of $\Tr \hat{\Gamma}^2$ for the particular
values $v=0.995$ and $v=0.999$ of the damping parameter, represented
respectively by crosses and circles.
The smallest individual values of  
$\Tr \hat{\Gamma}^2$ are insensitive to $v$ in this illustration. Physically,
these correspond to instances of the system being off resonance and
changing only slightly as $v$ makes the final steps towards unity. 
In the weak damping limit, where
resonances are very narrow, this accounts for most parameter values.
However, the minority of cases which are at or near resonance respond
very wildly to changing $v$ and become dramatically larger as $v$ 
is increased. These instances of resonant response bring the average
value of  $\Tr \hat{\Gamma}^2$ up to its predicted level (\ref{TrG}),
represented in Fig.~\ref{fig:Gfluc} by the horizontal lines.

Clearly then, the average fluctuations predicted in this paper present
only a partial characterisation of the system response in the weak
damping limit. A complete description demands a characterisation of
the distribution of values. This lies outside the scope of the current
discussion but will be reported in a future publication. The
calculations in this section do provide, however, a simple
characterisation of the variability of the response of the system
about the mean.

\section{Conclusion}
Ray tracing methods often provide the only feasible means of treating
wave propagation in the high-frequency limit. Recently
developed DEA methods have in particular provided an effective means of
implementing such phase-space transport 
in very large, complex structures. They provide an effective
description of the local averaged response of large systems to driving by
sources that are themselves also often complex and statistically 
characterised.
However, such methods used  in isolation miss important physical phenomena 
related to interference and to higher order effects such as diffraction.

The aim of this paper has been to establish and exploit an entirely 
wave-based analogue of this phase-space transport problem so that
wave effects such as multi-path interference can be
incorporated into large-scale simulations with complex and noisy
forcing. We have argued that an effective platform for the calculation 
of such wave effects can be built on the
relationship between correlation functions and Wigner functions 
that has been established in the contexts of quantum mechanics and optics.
Two-point correlation functions provide an effective means of
exploiting available information about spatial localisation and directionality
of waves radiated from a noisy, complex source. They have proven to be an
effective way of characterising 
EM emissions from electronic circuitry through direct 
measurement, see for example \cite{EMC15}.  Correlation
function propagators then provide a completely wave-based analogue of
phase-space transport approaches such as DEA.

Importantly, this allows us to provide a statistical description of 
fluctuations due to interference in the response of a forced wave
problem, using only information that is readily available from a
direct phase-space simulation. In particular, the global variance of the wave correlation
function can be described in terms of an autocorrelation function of
propagated phase-space densities. In essence this allows us to 
boot-strap phase-space transport simulations to predict fluctuation about
the mean of the response of the system, as well as the mean itself. 
The approach has been tested on simple quantum map models based on a
representation of wave transport by boundary transfer operators,
 but using a framework that we believe
will scale up effectively to much larger systems. 
Achieved results are relevant in the statistical characterisation of large vibro-acoustics 
and electromagnetic structures, including reverberation chambers operated at arbitrary frequencies.

\ack{We gratefully ackowledge support from from EPSRC under grant
  number EP/K019694/1, from the EU FP7 project MHiVec and from
  the EU Horizon 2020 network NEMF21.}

\appendix 
\section{The evolution of correlation functions} 
\label{app:A}

We will give here a derivation of Eq.\ (\ref{Wrho}) valid in the
semiclassical limit (see also \cite{MWH01}). 
Starting from Eq.\ (\ref{K}), we write 
$\hat{K_n} = \hat{T}^n \hat{\Gamma}_0 (\hat{T}^n)^\dagger$ in $x$
representation as 
\[K_n(x_1,x_2) = \int {\rm d}x'_1{\rm d}x'_2 \,T^n(x_1,x_1') K_0(x'_1,x'_2) (T^n)^\dagger(x'_2, x_2),
\]
where we set $K_0 = \Gamma_0$ for convenience. We employ the
short-wavelength approximation
\begin{equation} \label{BoT2}
T^n({x},{x'}) \approx \left(\frac{k}{2\pi \rmi}\right)^{(d-1)/2}
\sum_\alpha D_\alpha(x,x')\, \e^{\rmi k S_\alpha({x},{x'})}
\end{equation}
of the operator $\hat{T}^n$, 
 where the sum index $\alpha$ runs over all trajectories from $x'$ to $x$ that
encounter the SOS $n$ times, while other 
quantities are defined as in (\ref{BoT}) and (\ref{BoT1}) with
obvious modifications. For notational compactness, the phase factor 
$\mu$ is assumed to be absorbed into the definition of the amplitude 
here. 
Using the transformation
\begin{eqnarray}
x_1 = x + \frac{s}{2}; &\quad& x_1' = x' + \frac{s'}{2};\\
x_2 = x - \frac{s}{2}; &\quad& x_2' = x' - \frac{s'}{2};\nonumber
\end{eqnarray} 
and starting from  the semiclassical expression (\ref{BoT2}), we obtain
\begin{equation}
\begin{split}
K_n\left(x+\frac{s}{2},x - \frac{s}{2}\right) 
&\approx \left(\frac{k}{2 \pi}\right)^{d-1} \int {\rm d}x'{\rm d}s'
\, 
\sum_{\alpha\beta}\\
&D_\alpha\left(x+\frac{s}{2}, x'+\frac{s'}{2}\right) D_\beta^*\left(x-\frac{s}{2}, x'-\frac{s'}{2}\right)\\
& \times\e^{\rmi k (S_\alpha(x+s/2, x'+s'/2) - S_\beta(x-s/2, x'-s'/2))} K_0\left(x'+\frac{s'}{2},x'-\frac{s'}{2}\right),
\end{split}
\end{equation}
where $\alpha$ and $\beta$ label $n$-bounce orbits from
$x_1'$ to $x_1$ and from $x_2'$ to $x_2$, respectively.

At this point, we need to make two crucial assumptions about the
quantities of interest. First, we concentrate on the 
\textit{averaged} response, so that orbit combinations with
topologically distinct
$\alpha$ and $\beta$, which arrive with significant noncancelling phase 
differences, are washed out. We are then left just with the diagonal
contributions in which an orbit $\alpha$ coincides with an orbit $\beta$
up to a slight deformation. In the following we therefore set
$\alpha=\beta$.
Second, we assume that the source has a
sufficiently short correlation length that propagation can be
approximated using Taylor expansions of the surviving 
amplitude and phase contributions, truncated at the terms
\[
D_\alpha\left(x+\frac{s}{2}, x'+\frac{s'}{2}\right) 
D_\alpha^*\left(x-\frac{s}{2}, x'-\frac{s'}{2}\right) 
\approx |D_\alpha(x, x')|^2
\]
and 
\[
S_\alpha\left(x+\frac{s}{2}, x'+\frac{s'}{2}\right) - 
S_\alpha\left(x-\frac{s}{2}, x'-\frac{s'}{2}\right)
\approx p_\alpha(x,x') s - p_\alpha'(x,x') s'. 
\]
Here, $p_\alpha'(x,x') = -\partial S_\alpha(x,x')/\partial x'$ and
$p_\alpha(x,x') =\partial S(x,x')/\partial x$   \cite{LL}
are respectively the
initial and final momenta of the $n$-bounce trajectory 
from $x'$ to  $x$ labelled by $\alpha$. The condition 
that the correlation length is sufficiently short translates 
in the Wigner representation to the condition that 
the dependence on $p$ of the Wigner function is sufficiently 
slow. Note that even very smooth initial densities wrinkle under
chaotic evolution so that rapid oscillations develop as 
$n$ increased. For a fixed underlying dynamics, the truncated 
Taylor expansions above may therefore fail for moderately large 
$n$. By using the derived identities below for longer times we 
are therefore implicitly assuming that there is also sufficient
averaging over system parameters that very fine-scale classical
features are smoothed out in propagated densities.

With these assumptions we obtain
\[
\begin{split}
\left\langle K_n\left(x+\frac{s}{2},x - \frac{s}{2}\right)\right\rangle 
&\approx \left(\frac{k}{2 \pi}\right)^{d-1} \int {\rm d}x'{\rm d}s'
\, \sum_\alpha\\
&\left\langle|D_\alpha(x, x')|^2 \e^{\rmi k(p_\alpha s - p_\alpha's')} 
K_0\left(x'+\frac{s'}{2},x'-\frac{s'}{2}\right)\right\rangle.
\end{split}
\]
Using the definition of $D_\alpha$ given in (\ref{BoT1}), we may write
\[
\sum_\alpha \int {\rm d}x' |D_\alpha(x,x')|^2 \left\{\cdot\right\} = 
\left(\frac{k}{2 \pi}\right)^{d-1} \int {\rm d}p \left\{\cdot\right\},
\]
where on the left we sum over all orbits arriving at $x$, 
labelled by initial position $x'$ and topology $\alpha$.
On the right we reformulate the same sum as a simple 
integration over the momentum $p$ at arrival (in which 
there is no need for the label $\alpha$ since $x$ and $p$ uniquely 
determine the orbit topology). Using this reformulation of the sum we can 
write
\begin{equation} 
\begin{split}
\left\langle K_n\left(x+\frac{s}{2},x -
    \frac{s}{2}\right)\right\rangle 
&\approx 
\left(\frac{k}{2 \pi}\right)^{d-1} \int {\rm d}p\, {\rm d}s' 
\, \left\langle \e^{\rmi k (ps - p's')} \,
K_0\left(x'+\frac{s'}{2},x'-\frac{s'}{2}\right)\right\rangle
\\
& = \left(\frac{k}{2 \pi}\right)^{d-1} \int {\rm d}p \,\e^{\rmi k p
  s} \,
\left\langle W_{K_0} (x',p') \right\rangle\label{trafo}
\end{split}
\end{equation}
with $W_{K_0}(x',p')$ denoting the Wigner transform of $K_0$ as defined in (\ref{WDF}).
Note that $X' = (x',p')$ in (\ref{trafo}) is now a function of $X =
(x,p)$ 
through the relation $X = \varphi^n(X')$, with $\varphi$ defining the 
dynamics on the SOS (see Sec.\ \ref{sec:corr}).
 After applying the Wigner transformation on both sides of 
Eq.\ (\ref{trafo}) and evaluating the resulting $\delta$-function, we obtain

\begin{equation}
\left\langle W_{K_n}(X)\right\rangle = \left\langle 
W_{K_0}\left(\varphi^{-n}(X)\right)\right\rangle,
\end{equation}
which mirrors the action of the classical FP operator.

\section{Conventions for quantum maps} 
\label{app:B}
Here we summarise the quantum maps used in Secs.~\ref{sec:QKH} and \ref{numsec} to test correlation function
propagation. We use maps defined on a toral phase space with unit period
in each of the phase space coordinates $x$ and $p$. 

We begin with the classically-defined kicked map
defined by
\begin{eqnarray*}  
x &=& x' + G'(p')\\  
p &=& p' - F'(x),\qquad\mbox{(modulo 1)}
\end{eqnarray*}
where the primes on $F$ and $G$ denote differentiation.   This map can
be viewed as being the result  of 
using unit-time flow under the Hamiltonian
$
G(p)
$,
followed by unit-time flow under the Hamiltonian
$
F(x)
$.
The illustrations in Figs.~\ref{fig:kr:wig:a}-\ref{fig:kr:k1:a} assumed
$G'(p) = a\sin 2\pi p$ and $F'(x)=b\sin 2\pi x$ and $a= -2 b=2/\pi$,
in which case the map is a kicked Harper map, which will be the
nomenclature we use henceforth.
These maps provide generic examples with predominantly
chaotic dynamics mixed with small regular islands for the chosen parameter
values.

The corresponding quantum map is then defined on a Hilbert space of
dimension $N=1/(2\pi\hbar)$ such that
\[
\hat{U}_{\rm QKH} = \e^{-2\pi N\rmi F(\hat{x})}
\e^{-2\pi N\rmi G(\hat{p})}.
\]
The geometry of phase space is reflected in the detailed quantisation
of the position and momentum operators $\hat{x}$ and $\hat{p}$, and on
the boundary conditions imposed on quantum states. We can use a
position basis $\hat{x}\ket{x_i}=x_i\ket{x_i}$, with quantised positions
$x_i=(i+\alpha_x)/N$, whose index $i$ runs over
$i=0,\cdots,N-1$. Alternatively, we can use a momentum basis 
$\hat{p}\ket{p_i}=p_i\ket{p_i}$ on the same index set with quantised momenta
$p_i=(i+\alpha_p)/N$ and related to the position basis by
$\langle{x_l}|{p_j}\rangle=\e^{2\pi N \rmi x_lp_j}/\sqrt{N}$. The shifts
$\alpha_x$ and $\alpha_p$ are determined by the boundary conditions
satisfied by states under translation  across the torus. The direct
wavefunction $\psi(x_i)=\langle x_i|\psi\rangle$ satisfies
$\psi(x_i+1) = \e^{2\pi \rmi\alpha_p}\psi(x_i)$. The momentum
representation 
$\varphi(p_i)=\langle p_i|\psi\rangle$ satisfies
$\varphi(p_i+1) = \e^{-2\pi \rmi\alpha_x}\varphi(p_i)$. All of the maps
used in this paper use either $\alpha_x=\alpha_p=0$ or 
$\alpha_x=\alpha_p=1/2$ and, 
in particular, 
Figs.~\ref{fig:kr:wig:a}-\ref{fig:kr:k1:a} assumed
$\alpha_x=\alpha_p=0$.

Alternatively, cat maps and their quantisations provide examples of
fully chaotic, hyperbolic dynamical systems. Quantisations of the
simple cat map exhibit nongeneric degeneracies but these can be 
removed by perturbations which retain the fully chaotic dynamics. In
particular, we use a perturbed cat map
\[
\hat{U}_{\rm PC} = \hat{U}_{\rm C}\hat{U}_{\rm QKH},
\]
where $\hat{U}_{\rm QKH}$ is a quantum kicked Harper map and 
$\hat{U}_{\rm C}$, defined by
\[
\langle{x_j}|\hat{U}_C|x_l\rangle = \frac{1}{\sqrt{N}}\e^{2\pi N\rmi (x_j^2-x_jx_l+x_l^2/2)},
\]
quantises the (unperturbed) cat map
\begin{eqnarray*} 
x &=& x' + p'\\ 
p &=& x' + 2p' \qquad\mbox{(modulo 1).}
\end{eqnarray*} 
This quantisation is well-defined for even $N$ with
$\alpha_x=\alpha_p=0$ and for odd $N$ with 
$\alpha_x=\alpha_p=1/2$. The QKH map used for the numerical illustration in
Sec.~\ref{numsec} was also of this form with $G'(p)=a\sin
2\pi p$, $F'(x)=-b\sin 2\pi x$ and $a=b=0.1$. This choice completely eliminates 
symmetries in the combined map, while presenting a small enough
perturbation of the cat map that the dynamics is fully chaotic.

\end{document}